\documentclass[12pt]{article}
\usepackage{amssymb}
\usepackage{psfig}
\begin{document}
\begin{center}
\textbf{LIGHT  RAYS  AT  OPTICAL  BLACK  HOLES  IN  MOVING  MEDIA}\\

\bigskip
\bigskip
\bigskip

I. Brevik\footnote{E-mail address: iver.h.brevik@mtf.ntnu.no} and G. Halnes\footnote{E-mail address: geirhal@stud.fim.ntnu.no}\\

\bigskip
\bigskip

Division of Applied Mechanics,\\ Norwegian University of Science and Technology,\\
N-7491 Trondheim, Norway\\

\bigskip
\bigskip


\bigskip
\bigskip

Second revised version, September 2001
\end{center}

\bigskip

\begin{abstract}

Light experiences a non-uniformly moving medium as an effective gravitational field, endowed with an effective metric tensor $\tilde{g}^{\mu \nu}=\eta^{\mu \nu}+(n^2-1)u^\mu u^\nu$, $n$ being the refractive index and $u^\mu$ the four-velocity of the medium. Leonhardt and Piwnicki [Phys. Rev. A {\bf 60}, 4301 (1999)] argued that a flowing dielectric fluid of this kind can be used to generate an 'optical black hole'. In the Leonhardt-Piwnicki model, only a vortex flow was considered. It was later pointed out by Visser [Phys. Rev. Lett. {\bf 85}, 5252 (2000)] that in order to form a proper optical black hole containing an event horizon, it becomes necessary to add an inward radial velocity component to the vortex flow. In the present paper we undertake this task: we consider a full spiral flow, consisting of a vortex component plus a radially infalling component. Light propagates in such a dielectric medium in a way similar to that occurring around a rotating black hole. We calculate, and show graphically, the effective potential versus the radial distance from the vortex singularity, and show that the spiral flow can always capture light in both a positive, and a negative, inverse impact parameter interval. The existence of a genuine event horizon is found to depend on the strength of the radial flow, relative to the strength of the azimuthal flow. A limitation of our fluid model is that it is nondispersive. 

\end{abstract}
\newpage

\section{Introduction}

The classical topic of light propagation in moving dielectric media has received considerable interest in recent years - for a review of recent developments see, for instance, Ref. \cite{leonhardt00}. Much of the reason for the revitalization of this field of research is evidently the advent of the electromagnetically induced transparency (EIT) technique \cite{knight97}, whereby the light velocity in matter can be slowed down dramatically, to orders of 1-10 m/s \cite{hau99}. Under such circumstances it becomes possible to generate a fluid flow whose velocity is comparable to that of light in the fluid. The flow will then have an important influence upon the propagation of light.

Light waves experience a moving dielectric medium like an effective gravitational field. The effective metric, $\tilde{g}^{\mu \nu}=\eta^{\mu \nu}+(n^2-1)u^\mu u^\nu$, was used long time ago  to develop a relativistic model of a non-dispersive dielectric medium, and will be applied also in this paper. $n$ is the refractive index, and $u^\mu$ is the four-velocity of the medium.  The use of the effective metric allows us to apply the general-relativistic formalism to systems that normally are not identified with general relativity. Some earlier papers along these lines are Refs.[4-7]. Recent discussions on the general-relativistic analog models can be found at the web-address of the Rio de Janeiro Workshop in October 2000 \cite{workshop}. Since the velocity of light can be slowed down so dramatically, the flowing medium gives rise to the effect that light rays are moving along curved trajectories.

Central references in our present context are the papers of Leonhardt and Piwnicki \cite{{leonhardt99}, {leonhardt00a}}. These authors considered a dielectric fluid swirling around a straight linear vortex, arguing that such a system can be used to generate an 'optical black hole'. Subsequent remarks of Visser \cite{visser00} made it clear, however, that the interpretation of the Leonhardt-Piwnicki model in its original form as an optical black hole is not feasible: in order to obtain a black hole endowed with an event horizon one has to add an {\it inward} radial component to the vortex flow (cf. also the rebuttal \cite{leonhardt00b}). This brings us to the subject of the present paper: We will develop the optical black hole theory based upon a general spiral flow model, consisting of a vortex component plus a radially infalling component. We introduce an "effective potential", constructed as the appropriate generalization of the conventional effective potential outside a Schwarzschild black hole \cite{misner73}. Plotting the variation of the effective potential versus radial distance $r$ from the vortex singularity, we get a powerful tool for predicting particle and light trajectories. Our explicit calculations prove that Visser's remarks \cite{visser00} (which are based upon general considerations) are right. We find that the spiral flow can always capture light within a positive, and a negative, interval for the inverse impact parameter. If the azimuthal velocity component is increased, the parameter interval for light being able to escape from the system is also increased. This reminds us of gravitational black holes in that way that it is known that a rotating black hole captures light less efficiently than a static one of same mass.

One important limitation of our theory should be emphasized: we consider {\it non-dispersive} media only. That means, we are unable to cope with the extremely dispersion-sensitive EIT effect \cite{hau99}. Some more comments should be given, concerning the physical background for this statement. Although the EIT technique is extremely effective - the quantum interference allows pulse transmission through an atomic cloud which would otherwise have a transmission coefficient only of order $e^{-110}$ - it is notable that the effect is also extremely dispersive.  The steep dispersive profile is shown by Fig.~2 in the article of Hau {\it et al.} \cite{hau99}. According to the figure the high refractive index prevails only over a frequency interval of about $ \Delta \nu = 2$ MHz around the central frequency $\nu=0.5\times 10^{15}$ Hz (corresponding to a wavelength of  $\lambda=589$ nm). When the light propagates in the direction of motion of the fluid, it follows from a consideration of the four-momentum $k_\mu$ of the photon that the fluid velocity $v$ gives rise to a Doppler shift of $\Delta \nu= \nu n v/c$, or $v=(c/n)(\Delta \nu/\nu) = 120/n$ cm/s. Now, in the atomic vapour $n$ is practically unity, so we see that already  a moderate fluid velocity of about 100 cm/s is sufficient for the Doppler effect to bring the frequency in the local rest inertial frame of the fluid outside the EIT-generated high-$n$ band. The Doppler argument assumes for simplicity that $n$ is a constant, but this is not disturbing for our estimate: both is $n$ in the undisturbed vapour close to unity, as mentioned, and moreover the refractive index-disturbance caused by the laser beams is very small, $\Delta n = 1.8\times 10^{-6}$. In practice, dispersive effects thus play an important role; the Doppler detuning will exceed the EIT window, and the present theory is not directly applicable to slow light. We think, however, that the nondispersive theory is of interest in its own right.

In the next section we summarize the basic properties of the phenomenological electromagnetic theory in a dielectric medium, emphasizing the formal analogy to general relativity, and also the connection between this kind of theory and the transformation procedure worked out some years ago \cite{brevik70}, from which one can map the electromagnetic field in a vacuum to that in a medium.  Then, in Sec. 3 we consider the effective metric for the two components of the spiral flow. In Sec. 4, before embarking on the rather complex theory of the general spiral case, we define the "effective potential", which is a new concept in this context, and apply it to the two basic flow components separately (the pure sink, and the pure vortex). Figures 2 and 3 show the "effective potentials" for these two cases. The figures are useful, in that they show which regions are inaccessible for photon trajectories. In Sec. 5, which is the main section, we work out the "effective potential" for the full spiral flow, and introduce various concepts: the event horizon, the apparent horizon, the ergo region, and the static limit. Various kinds of photon trajectories are discussed in relation to Fig. 4. As far as we know, this kind of analysis has not been given before, for the general composite flow. A summary is given in Sec. 6, along with a closer discussion on the relationship to a Schwarzschild black hole.

We put $c=1$, adopt the Minkowski metric convention $\eta_{\mu \nu}=(1,-1,-1,-1)$, and follow the basic formulation of Leonhardt and Piwnicki \cite{leonhardt99}. We mention in passing that some other related recent treatises can be found in Refs. [14-19].

\section{Basic formalism}

We begin by giving a brief account of the basic formalism, letting in this section the four-velocity $u^\mu$ of the fluid be unspecified (thus not necessarily being the sum of vertex flow and radial flow). With the above metric, $u^\mu u_\mu =1$. If $U^i$ are the components of the three-dimensional velocity vector ${\bf U} $, i.e.,  $U^i U_i= {\bf U}^2$, we have thus  
\begin{equation}
u^\mu=\gamma(1, U^i),
\end{equation}
\label{1}
where $\gamma$ is the Lorentz factor, $\gamma=(1-{\bf U}^2)^{-1/2}$.

It ought to be emphasized that we are applying a kind of hybrid interpretation to the system. First, as far as the fluid flow velocity $u^\mu$ and the wave four-vector $k_\mu$ are concerned, we consider the system as an ordinary special-relativistic system, in which covariant and contravariant indices are moved by means of the Minkowski metric $\eta_{\mu\nu}$. Second, when considering the wave propagation through the moving flow, we make use of the effective metric, which will be designated by $\tilde{g}_{\mu\nu}$. The background for introducing this kind of metric can be seen as follows. Starting from Maxwell's equations for a pure radiation field in a nondispersive medium moving with special-relativistic velocity $u^\mu$ we obtain, by defining the four-potential $A_\mu$ from the basic field tensor $F_{\mu\nu}$ via the relationship $F_{\mu\nu}=\partial_\mu A_\nu -\partial_\nu A_\mu $, the field equation \cite{brevik70}
\begin{equation}
[ {\square} +\kappa (\partial.u)^2]A_\mu=0
\end{equation}
\label{2}
($\square \equiv \partial_\mu \partial^\mu, ~~\partial.u \equiv \partial_\mu u^\mu)$, where we have defined $\kappa $ as
\begin{equation}
\kappa=n^2-1.
\end{equation}
\label{3} 
In momentum space, Eq.~(2) implies
\begin{equation}
k^2+\kappa (k\cdot u)^2=0
\end{equation}
\label{4}
($k^2 \equiv k^\mu k_\mu$). Since the indices of the photon four-momentum  $k^\mu$, as mentioned,  can be lowered and raised by the Minkowski metric $\eta_{\mu\nu}$, we can write Eq.~(4) as 
\begin{equation}
\tilde{g}^{\mu\nu}k_\mu k_\nu=0,
\end{equation}
\label{5}
with $\tilde{g}^{\mu\nu}$ being the effective metric
\begin{equation}
\tilde{g}^{\mu\nu}=\eta^{\mu\nu}+\kappa u^\mu u^\nu.
\end{equation}
\label{6}
It is the appearance of Eq.~(5) that makes it possible to borrow terminology from general relativity and consider the propagation of light waves as taking  place in a fictitious or 'effective' curved space whose metric is given by Eq.~(6).

In the limiting case of Hamilton's geometrical optics, which is adopted here, the electromagnetic field tensor is modelled by the eikonal ansatz \cite{landau75}. It implies that the photon four-momentum is essentially the four-divergence of the eikonal $S$, i.e., $k_\mu = -\partial_\mu S$. Accordingly, it becomes natural to write the wave equation in such a way that it involves the {\it covariant} components of the wave vector, as in Eq.~(5). 

We mention in passing the following point. In Ref.~\cite{brevik70}, we developed a transformation procedure which maps the electrodynamic theory for a vacuum field directly into the phenomenological nondispersive theory for a medium field. This mapping procedure is actually quite powerful, especially in connection with quantization of the medium field. Consistent with the use of the canonical (space-like) four-momentum $k_{\mu}$ in the medium, the quantization is done in accordance with the nonsymmetric, divergence-free Minkowski energy-momentum tensor. The central building block in this kind of theory is the matrix $b^{\mu\nu}$, defined by
\begin{equation}
b^{\mu\nu}=\eta^{\mu\nu} + (n-1)u^\mu u^\nu,
\end{equation}
\label{7}
satisfying the relationship ($p$ an integer):
\begin{equation}
(b^p)^{\mu\nu}=\eta^{\mu\nu}+(n^p-1)u^\mu u^\nu.
\end{equation}
\label{8}
The effective metric  in Eq.~(6) is thus nothing else than the second power of the basic matrix:  $\tilde{g}^{\mu\nu}  \equiv (b^2)^{\mu\nu}$.

\section{The two flow components}

We now turn to the situation in which the flow, as mentioned above, is composed of a vortex component and a radial component. The three-dimensional fluid velocity ${\bf U}$ can thus be written as
\begin{equation}
{\bf U}= U^{\hat{r}}{\bf e}_{\hat r}+ U^{\hat \phi}(r) {\bf e}_{\hat \phi},
\end{equation}
\label{9}
in an orthogonal basis (designated by hats) $ \{{\bf e}_{\hat i} \}
 = \{ {\bf e}_{\hat r}, 
\bf{e}_{\hat \phi}, {\bf e}_{\hat z} \}$. We write the azimuthal component in the form  $U^{\hat \phi}=\Omega/r$ with $\Omega$ a constant,  $r$ being the radial distance from the singular vortex (situated along the $z$ axis). In a coordinate basis,  $ \{ {\bf e}_i\}=\{ \partial/\partial r, \partial/\partial \phi, \partial/\partial z \}$, this corresponds to $U^\phi =\Omega/r^2$. The relativistic correction is introduced \cite{leonhardt99} via the Lorentz factor $\gamma$,
 so that
 $U^\phi \rightarrow  \Omega/(\gamma r^2)$.
 As for the radial flow, we start from the nonrelativistic  standard form for a singular line source, $U^{\hat r} = \Gamma/r$, $\Gamma$ being a constant that is negative when, as assumed here, the source is actually a sink. (There is for the radial component no difference between the orthogonal basis and the coordinate basis, i.e., $U^{\hat r} = U^r.$) Again including the relativistic correction, we have 
$U^r \rightarrow \Gamma /(\gamma r)$.
 Altogether, the three-dimensional fluid velocity ${\bf U}$ in coordinate basis becomes decomposed as
\begin{equation}
{\bf U}=(U^r, U^\phi, U^z)= ( \frac{\Gamma}{\gamma r}, \frac{\Omega}{\gamma r^2}, 0 ),
\end{equation}
\label{10}
with Lorentz factor
\begin{equation}
\gamma= \sqrt{1+\frac{\Omega^2+\Gamma^2}{r^2}}.
\end{equation}
\label{11}
From Eq.~(1) we obtain the fluid's four-velocity:
\begin{equation}
u^\mu = (u^0, u^r, u^\phi, u^z)=\left( \gamma, \frac{\Gamma}{r}, \frac{\Omega}{r^2}, 0 \right).
\end{equation}
\label{12}
There are two nontrivial Killing vectors in the problem, viz. $\xi_{(t)}=\partial/\partial t$ and $\xi_{(\phi)}=\partial/\partial \phi$. The four-vector products $k\cdot \xi_{(t)}$ and $k \cdot \xi_{(\phi)}$ are constant along the photon trajectory, meaning that the components $ k_0$ and $k_\phi$ are constants. We will designate them respectively by $E$ and $-L$:
\begin{equation}
k_0=E,~~~~k_\phi=-L.
\end{equation}
\label{13}
The physical meaning of these quantities can be seen as follows.  Consider the line element for light at spatial infinity:
\begin{equation}
ds^2= \tilde{g}_{\mu\nu}dx^\mu dx^\nu \rightarrow \frac{1}{n^2}dt^2-dr^2-r^2d\phi^2-dz^2.
\end{equation}
\label{14} 
In an orthonormal basis $\{\omega^{\hat \mu }\}$ at $r \rightarrow \infty$  we thus have that the zeroth component becomes $\omega^{\hat 0} = dt/n$. The local photon energy $E_{local}$ at infinity is the contraction of the four-momentum $\tilde{k}^\mu$ with $\omega^{\hat 0}$:
\begin{equation}
E_{local}=\langle \omega^{\hat 0},\tilde{k} \rangle = \frac{\tilde{k}^0 }{n}= n k_0,
\end{equation}
\label{15}
since $\tilde{g}^{00} = n^2$. Thus
\begin{equation}
E=\frac{E_{local}}{n}.
\end{equation}
\label{16}
Since $E$ and $E_{local}$ differ only through a constant factor $n$, we can use these two concepts  interchangeably.

As for the azimuthal part, the interpretation of $L$ can be taken over directly from the theory of an ordinary black hole \cite{misner73}. The spatial components of the metric  at infinity are identical to those of the Minkowski metric, for the Schwarzschild metric as well as for the present case with  effective metric $\tilde{g}^{\mu\nu}$. The quantity $L$ is the angular momentum (for an ordinary black hole referring to equatorial orbits). The minus sign in front of the second of  Eqs.~(13) is due to our metric convention.

The quantity $b$, defined by
\begin{equation}
b=\frac{L}{E},
\end{equation}
\label{17}
will henceforth be referred to as the impact parameter (at infinity). (This is not in perfect agreement with the usual definition of the impact parameter.) As discussed by Leonhardt and Piwnicki \cite{leonhardt99} the canonical angular momentum $L$ differs from the {\it kinetic} angular momentum by a constant, due to the optical Aharonov-Bohm effect.

The light paths are defined by the wave four-vector $\tilde{k}^\mu$ tracing out zero geodesics of $\tilde{g}^{\mu\nu}$. Introducing, as usual \cite{misner73}, an affine parameter $\lambda$, we can parametrize the light path through the relation
\begin{equation}
\tilde{k}^{\mu}=\frac{dx^\mu}{d\lambda}.
\end{equation}
\label{18}
Of main interest for us is the $r-$ component:
\begin{equation}
\tilde{k}^r=\frac{dr}{d\lambda}.
\end{equation}
\label{19}
It is to be noted that when $dr/d\lambda$ is zero the photon, with its radial momentum equal to $\hat{k}^r$, has a {\it turning point}.

For reference purposes we write down the components of the effective metric for the full spiral flow. From Eqs.~(6), (10), and (11) we have
\[ \tilde{g}^{00}= 1+ \kappa \left( 1+\frac{\Omega^2+\Gamma^2}{r^2} \right), \]
\[ \tilde{g}^{0r}= \frac{\kappa \Gamma}{r}\sqrt{1+\frac{\Omega^2+\Gamma^2}{r^2}},  \]
\[ \tilde{g}^{0 \phi}=\frac{ \kappa \Omega}{r^2}\sqrt{1+\frac{\Omega^2+\Gamma^2}{r^2}}, \]
\begin{equation}
\tilde{g}^{rr}= -1+\frac{\kappa \Gamma^2}{r^2},
\end{equation}
\label{20}
\[ \tilde{g}^{r \phi}=\frac{\kappa \Gamma \Omega}{r^3}, \]
\[ \tilde{g}^{\phi \phi}= -\frac{1}{r^2}+\frac{\kappa \Omega^2}{r^4}, \]
\[ \tilde{g}^{zz}= -1. \]
We will need also the covariant components:
\[ \tilde{g}_{00}=1-\frac{\kappa}{n^2}\left( 1+\frac{\Omega^2+\Gamma^2}{r^2} \right), \]
\[ \tilde{g}_{0r}=\frac{\kappa \Gamma}{n^2 r}\sqrt{1+\frac{\Omega^2+\Gamma^2}{r^2}}, \]
\[ \tilde{g}_{0 \phi}=\frac{\kappa \Omega}{n^2}\sqrt{1+\frac{\Omega^2+\Gamma^2}{r^2}}, \]
\begin{equation}
\tilde{g}_{rr}=-1-\frac{\kappa \Gamma^2}{n^2r^2}, 
\end{equation}
\label{21}
\[ \tilde{g}_{r \phi}=-\frac{\kappa \Omega \Gamma}{n^2r}, \]
\[ \tilde{g}_{\phi \phi}=-r^2-\frac{\kappa \Omega^2}{n^2}, \]
\[ \tilde{g}_{zz}=-1. \]
These expressions satisfy the contraction conditions $\tilde{g}^{\mu \alpha} \tilde{g}_{\mu\beta}=\delta^\alpha_\beta$.

It ought to be pointed out here that we have constructed the velocity of the composite flow, Eqs.~(10)-(12), as an ad hoc generalization of the flow adopted by Leonhardt and Piwnicki \cite{{leonhardt99},{leonhardt00a}}. Since the four-velocity $u^\mu$ has to satisfy the relativistic Euler equation, the value of the pressure $p$ has to adjust itself accordingly. We do not consider the pressure here. It should also be mentioned that the values of the metric components given by Eqs.~(20) and (21) are different from those used by Linet \cite{linet00}.

\section{"Effective potential" for the spiral flow. Application to the special cases}

The Hamilton-Jacobi equation, Eq.~(5), will now be solved for the general planar centrally symmetric flow. This spiral flow model of course gives associations to a Schwarzschild black hole. We will show that the system indeed gives rise to effects similar to those occurring near rotating black holes. Thus the concepts of an event horizon, and an ergosphere, can be introduced in a natural way, and we will also show how qualitative features of light trajectories in the flow are obtained by plotting the "effective potential" for light. First, the expression for the "effective potential" has to be derived.

\subsection{The "effective potential" for the spiral flow}

Employing the effective metric picture, we will use $\tilde{g}^{\mu \nu}$ to raise indices of the covariant wave vector $k_\mu$. Thus the contravariant effective wave vector $\tilde{k}^\mu$ will be defined as
\begin{equation}
\tilde{k}^\mu= \tilde{g}^{\mu \nu} k_\nu,
\end{equation}
\label{22}
the inverse equation being $k_\mu=\tilde{g}_{\mu\nu}\tilde{k}^\nu$. (Note that the covariant vector $k_\mu$ retains its usual special-relativistic meaning.) In order to solve Eq.~(5), one needs to find all the components of $k_\mu$  expressed in terms of known quantities. Since three components of $k_\mu$ are constants, $k_0=E,~k_\phi =-L, ~k_z=0$, it remains to find the component $k_r$. To this end we first express $k_r$ as
\begin{equation}
k_r= \tilde{g}_{r0} \tilde{k}^0+\tilde{g}_{rr}\tilde{k}^{r}+\tilde{g}_{r\phi} \tilde{k}^\phi,
\end{equation}
\label{23}
and thereafter insert $\tilde{k}^0=\tilde{g}^{0\nu}k_\nu ,~\tilde{k}^\phi=\tilde{g}^{\phi \nu} k_\nu$ to convert Eq.~(23) to an expression in which all components except from $k_r$ are known. Taking into account the constants above, we can then write $k_r$ as
\begin{equation}
k_r=\frac{A}{C}E-\frac{B}{C}L+\frac{\tilde{g}_{rr}}{C}\frac{dr}{d\lambda},
\end{equation}
\label{24}
where
\[ A= \tilde{g}_{0r}\tilde{g}^{00}+\tilde{g}_{\phi r}\tilde{g}^{0\phi}, \]
\begin{equation}
B=\tilde{g}_{0r}\tilde{g}^{0\phi}+\tilde{g}_{\phi r}\tilde{g}^{\phi \phi},
\end{equation}
\label{25}
\[ C=1-\tilde{g}_{0r}\tilde{g}^{0r}-\tilde{g}_{\phi r}\tilde{g}^{\phi r}. \]
At the turning point of a trajectory $dr/d\lambda =0$, so that the last term in Eq.~(24) vanishes. Turning points will be a matter of main interest below. It is convenient, therefore, to introduce a new symbol $\hat{k}_r$ denoting the value of $k_r$ at just the turning point. It is given by
\begin{equation}
\hat{k}_r=\frac{A}{C}E-\frac{B}{C}L.
\end{equation}
\label{26}
All components of $k_\mu$ in the Hamilton-Jacobi equation are thereby known. If we for simplicity consider the solution of this equation accepting the condition (26), the Hamilton-Jacobi equation determines which turning-point value of $r$ is compatible with given input values of $E$ and $L$.  To emphasize notationally this particular case, we give an extra 'hat' to the symbols, so that $E \rightarrow \hat{E},~L \rightarrow \hat{L}$. Whereas for given values of $\hat{E}$ and $\hat{L}$ the equation determines for which $r$ the turning of the photon trajectory takes place, it should be noted that at the same $r$ other photons, corresponding to different values of $E$ and $L$ are also allowed to exist, but they do not have turning points at this position.

As will be discussed in more detail below, the turning point solutions give the effective potential for the system. The Hamilton-Jacobi equation is a quadratic equation in $\hat{E}$ (or alternatively in $\hat{L}$); it is convenient to work instead in terms of the ratio between these quantities, i.e., $\hat{b}^{-1}=\hat{E}/{\hat{L}}$, which is the inverse impact parameter. We write the Hamilton-Jacobi equation in the form
\begin{equation}
\mathcal{P} ~\hat{b}^{-2}+\mathcal{Q}~ \hat{b}^{-1}+\mathcal{R}= 0,
\end{equation}
\label{27}
with known coefficients $\mathcal{P},~\mathcal{Q},~\mathcal{R}$. This equation has two sets of solutions, $\hat{b}^{-1}_{+} (r)$ and $\hat{b}^{-1}_{-} (r)$, which in the following will be called the "effective potentials".

Some remarks are called for here, since we are using this concept in a somewhat more restricted sense than what is usually the case. Recall the conventional equation determining the azimuthal angle $\varphi$:
\[ \phi -\phi_0=\int \frac{(L/r^2)dr}{\sqrt{2m[E_0-V_{eff}(r)]}},~\]
$m$ and $E_0$ being constants. In this case the $r$-dependent function $V_{eff}(r)$ is naturally called an effective potential. For example, trajectories around a Schwarzschild black holes can be expressed in this form. The physical meaning of our "effective potential" is simply that it corresponds to a curve from which the turning points for photons can be read off. In other words, whereas in Ref. \cite{misner73}, for instance, the effective potential is part of the equation of motion, regardless of whether the trajectory has a turning point or not, in our model the "effective potential" is found by solving the equation of motion precisely at a turning point. Our "potential" moreover does not have a nonrelativistic analogy. 

The expressions for $\mathcal{P},~\mathcal{Q},~\mathcal{R} $ are at first sight complicated, but after a  lengthy algebra making use of the metric components (20), (21) as well as Eqs.~(25), (26) we arrive at the following relatively simple expressions:
\[ \mathcal{P}=\frac{\kappa \Omega^2 +n^2 r^2}{r^2-\kappa \Gamma^2}, \]
\begin{equation}
\mathcal{Q}=\frac{-2\kappa \Omega}{r^2-\kappa \Gamma^2}\sqrt{1+\frac{\Omega^2+\Gamma^2}{r^2}}, 
\end{equation}
\label{28}
\[ \mathcal{R}=\frac{\kappa^2(\Omega^2+\Gamma^2)-r^2}{(\kappa \Gamma^2-r^2)r^2}. \]
The solutions of Eq.~(27) now become, when we make use of the expressions (28), 
\begin{equation}
\hat{b}_{\pm}^{-1}=\frac{\kappa \Omega}{\kappa \Omega^2+n^2 r^2}\sqrt{ 1+\frac{\Omega^2+\Gamma^2}{r^2}} \pm
\frac{\sqrt{n^2(r^2-\kappa \Gamma^2)}}{\kappa \Omega^2+n^2r^2}.
\end{equation}
\label{29}

\subsection{Inaccessible regions}

In order to justify our interpretation of the expression (29) as an "effective potential", we shall now show how it acts an an impenetrable barrier. Let us assume temporarily that $dr/d\lambda$ is nonzero. Then $k_r$, instead of $\hat{k}_r$ given by Eq.~(26), has to be inserted in Eq.~(5). The Hamilton-Jacobi equation accordingly becomes
\begin{eqnarray}
\mathcal{P}~b^{-2}+\mathcal{Q}~b^{-1} +\mathcal{R}
 &=&  \left[ -2\tilde{g}^{0r} E+2\tilde{g}^{rr} \left( \frac{BL-AE}{C} \right)+2\tilde{g}^{r\phi} L\right] 
\frac{\tilde{g}_{rr}}{C}\frac{dr}{d\lambda} \nonumber \\
 &-& \tilde{g}^{rr} \frac{(\tilde{g}_{rr})^2}{C^2} \left( \frac{dr}{d\lambda}\right)^2,
\end{eqnarray}
\label{30}
with the same factors (25) and (28) as before. Here $b^{-1}$ has taken the place of $\hat{b^{-1}}$, since turning points solutions are now not searched for.

Consider the left hand side of Eq.~(30). From Eq.~(28) it follows that if
\begin{equation}
r^2 > \kappa \Gamma^2,
\end{equation}
\label{31}
then $\mathcal{P} >0$. If the left hand side is plotted versus $b^{-1}$ the curve gets a positive curvature, and crosses the abscissa axis at the turning point solutions, $\hat{b}_{\pm}^{-1}$, as indicated in Fig.~1.

Consider next the right hand side of Eq.~(30). The first term turns out to be zero, as it must be, since the right hand side can not depend on the sign of $dr/d\lambda$ while the left does not. Thus, Eq.~(30) becomes
\begin{equation}
\mathcal{P}~b^{-2}+\mathcal{Q}~b^{-1} +\mathcal{R}=-\tilde{g}^{rr}\frac{(\tilde{g}_{rr})^2}{C^2}
\left( \frac{dr}{d\lambda} \right)^2.
\end{equation}
\label{32}
From Eq.~(20) it follows that $\tilde{g}^{rr}$ is negative in the outer region defined by Eq.~(31), so that the right hand side of Eq.~(32) is positive in this region. The condition that both sides of Eq.~(32) must be positive, makes us thus conclude that the region between $\hat{b}_{-}^{-1}$ and $\hat{b}_{+}^{-1}$ in Fig.~1 is inaccessible. The solution of the Hamilton-Jacobi equation (32) accordingly contains a potential barrier.  

\begin{figure}[htb]

\centerline{\mbox{\psfig{file=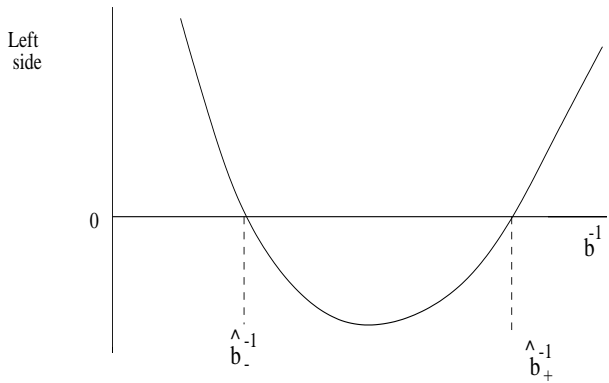,width=8cm,height=5cm,angle=0}}}
\caption{Qualitative sketch of the left hand side of Eq.(32) versus $b^{-1}$, when $r^2>\kappa\Gamma^2$.}
\label{opptur}
\end{figure}

Let us now analyse the inner region,
\begin{equation}
r^2 < \kappa \Gamma^2.
\end{equation}
\label{33}
From Eq.~(29) it follows that there are no real solutions for turning points in this region; the left hand side of Eq.~(30) has no zeros. The inequality (33) implies that $\cal{P}$ is negative, so that the curve in Fig.~1 is to be   replaced by a curve (not shown here) which is concave downwards and which has no intersection point with the abscissa axis. Since  $\tilde{g}^{rr}$ is now positive, the left hand side of Eq.~(30) is negative for any value of the impact parameter.  Any photon, if it comes that far, is allowed to exist for any value of $r$ in the inner region. There is no potential barrier, so that a photon approaching the singularity will ultimately get absorbed there.

\subsection{ Application to the perfect sink flow}

Before embarking on an analysis of the full spiral flow, it is worthwhile to consider the two special cases first.

The simplest special case is that of a perfect sink, corresponding to $\Omega =0,~\Gamma <0$. According to  Eq.~(29) the effective potential is now
\begin{equation}
\hat{b}_{\pm}^{-1}=\pm \frac{\sqrt{r^2-\kappa \Gamma^2}}{nr^2}.
\end{equation}
\label{34}
The "effective potential" is shown qualitatively in Fig.~2. The symmetry of the curve about the abscissa axis reflects the physical property that the turning point of a photon with angular momentum $L$ must occur at the same value of $r$  as the turning point of a photon with angular momentum $-L$.

\begin{figure}[htb]
\centerline{\mbox{\psfig{file=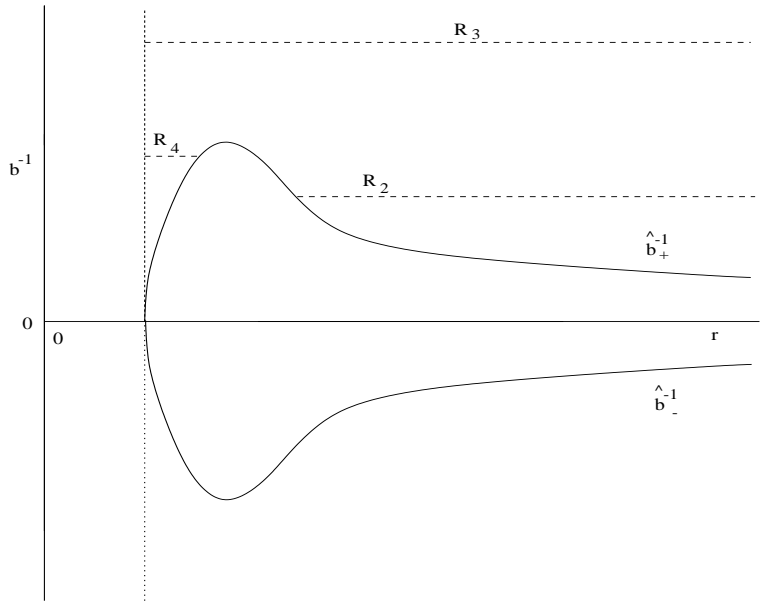,width=10cm,height=8cm,angle=0}}}
\caption{Perfect sink: effective potential, $\hat{b}^{-1}_{\pm}=\hat{E}/\hat{L}$, versus r. Broken lines, $R_2$, $R_3$ and $R_4$ refer to the three possible photon-trajectories. }
\label{sinkfig}
\end{figure}

The present "effective potential" is essentially the same as the effective potential for light in Schwarzschild geometry \cite{misner73}. At the position determined by
\begin{equation}
r^2=\kappa \Gamma^2,
\end{equation}
\label{35}
the effective potential becomes zero. This point is interpreted as the {\it event horizon} of the system. At smaller $r$ there is not, for any impact parameter, a turning point. 

The fluid's three-dimensional velocity in coordinate basis is, according to Eq.~(10),
\begin{equation}
{\bf U}=\left( \frac{\Gamma}{\gamma r}, 0, 0 \right),~~\rm{with}~~ \gamma=\sqrt{1+\frac{\Gamma^2}{r^2}}.
\end{equation}
\label{36}
Now, the event horizon would from a physical viewpoint be naturally defined as the position where the velocity of light relative to the medium, $1/n$, is equal to the inward radial fluid velocity, $\Gamma/\gamma r$. This leads to the condition $1/n^2=\Gamma^2/\gamma^2 r^2$ which, upon insertion of the Lorentz factor from Eq.~(36), is just the same as Eq.~(35). This shows the consistency of identifying the zero point of the effective potential with the event horizon .

The effective potential curve for a perfect sink is essentially the same as for light outside a Schwarzschild black hole. Whereas a conventional black hole can be visualized as a sink of space-time, the present 'optical black hole' can analogously be visualized as a sink of the moving medium, which at the same time serves as the artificial 'space-time'.

The three horizontal broken lines marked by $R_2, R_3, R_4$ in Fig.~2 refer to the fact that there are three different types of motion for photons in the radial flow. It is here first of all appropriate to recall that in Schwarzschild geometry there are  actually {\it four} types of motion possible when the orbiting particle has nonzero mass \cite{novikov89}. The first type, which we would naturally denote by $R_1$, refer to the 'pit' in the effective potential that permits stable orbits to exist around the black hole. Now, there are no stable orbits for light, so that Region $R_1$ does not exist, neither in Schwarzschild geometry nor in 'perfect sink' geometry. 

The other three regions exist for light, in Schwarzscild geometry as well as in the present case:

Light can approach the centre of the optical black hole with an inverse impact parameter that lies above the "effective potential". If $b^{-1}$ at some radius is equal to the "effective potential" $\hat{b}_+^{-1}$, this will be the turning point for the trajectory. After turning, the light goes outwards and escapes to infinity. This is illustrated as Region 2. 

If the photon has an inverse impact parameter that lies above the maximum of the "effective potential", it will be captured by the optical black hole. This is shown as Region 3. The region also covers the case where a photon, emitted somewhere outside the event horizon, escapes to infinity.

Region 4 is the region for a photon which is emitted somewhere near the event horizon and eventually is lost to the inside region. It may at first propagate in the outward direction, but 'hits' the potential curve to the left of the maximum point, and turns back into the centre of the optical black hole. 

Finally, we comment briefly on the unstable photon orbits that may exist temporarily  in the pure sink flow. Differentiating Eq.~(34) we see that the extrema of the "effective potential" are lying at 
\begin{equation}
r^2= 2\kappa \Gamma^2.
\end{equation}
\label{37}
The two extrema are lying at the same position $r$ (which is physically obvious since they describe photons travelling in opposite directions past the optical black hole), and they correspond to unstable orbits. The corresponding extrema follow from Eq.~(34) as
\begin{equation}
\hat{b}_{\pm}^{-1}=\pm \frac{1}{2n\sqrt{\kappa}\, \Gamma}.
\end{equation}
\label{38}
Evidently  such orbits will never be permanent; a photon having impact parameter very close to the critical one given by  Eq.~(38) will spiral several times around the centre before it either escapes to infinity or else becomes lost inside the event horizon.

\subsection{Application to the perfect vortex flow}

This is the case studied by Leonhardt and Piwnicki \cite{leonhardt99, leonhardt00a}. It corresponds to $\Gamma=0$. We will assume a positive vortex, so that $\Omega > 0$. From Eq.~(29) we get the "effective potential":
\begin{equation}
\tilde{b}_{\pm}^{-1}=\frac{\kappa \Omega \sqrt{1+\Omega^2/r^2} \pm nr}{\kappa \Omega^2+n^2 r^2}.
\end{equation}
\label{39}
A qualitative plot of the effective potential is shown in Fig.~3. The rotation-induced asymmetry is obvious. The upper curve corresponds to the positive sign in Eq.~(39). When $r$ approaches zero, both solutions diverge. 
As in the general case, the region between the two solutions is inaccessible.

When $\Omega$ is positive, the motion is in the positive $\phi$ direction. Positive rotation of the fluid corresponds to positive angular momentum, so that photons with $L>0$ move along the medium whereas photons with $L<0$ are fighting against the flow. This is the physical reason why photon trajectories corresponding to $L$ and $-L$ are non-symmetric in Fig.~3.

\begin{figure}[htb]
\centerline{\mbox{\psfig{file=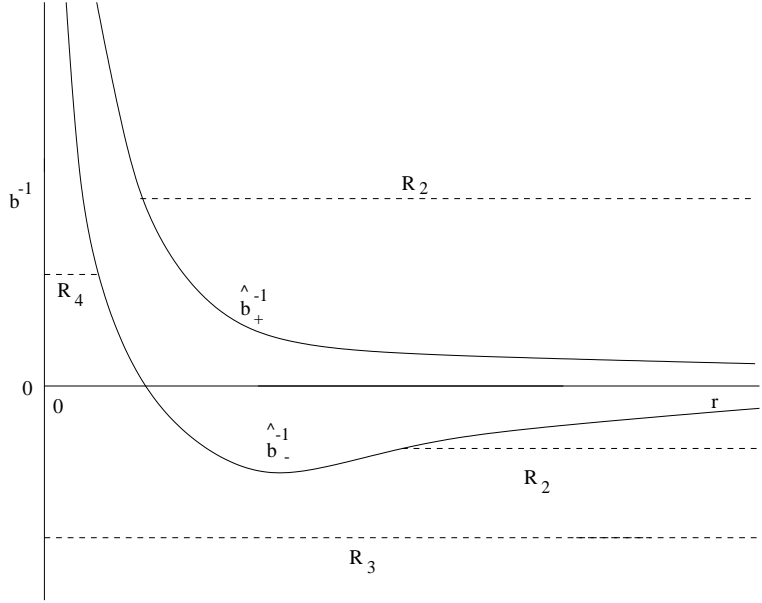,width=10cm,height=8cm,angle=0}}}
\caption{Perfect vortex: effective potential, $\hat{b}^{-1}_{\pm}=\hat{E}/\hat{L}$, versus r. Broken lines, $R_2$, $R_3$ and $R_4$ refer to the three possible photon-trajectories. }
\label{vortfig}
\end{figure}

For later use, let us introduce the {\it ergo region} concept. It will be showed below that the boundary of this region is at the point where the negative solution $\hat{b}_-^{-1}$  for the "effective potential" in Fig.~1 coincides with the origin of the $b^{-1}$ axis:
\begin{equation}
\hat{b}_-^{-1}(r)=0.
\end{equation}
\label{40}
From Eq.~(39) this means
\begin{equation}
r=\sqrt{\kappa}\,\Omega.
\end{equation}
\label{41}  
As from Eq.~(10) we have, for vortex flow,
\begin{equation}
{\bf U}=\left( 0, \frac{\Omega}{\gamma r^2},0 \right),
\end{equation}
\label{42}
with
\begin{equation}
\gamma=\sqrt{1+\frac{\Omega^2}{r^2}}
\label{43}
\end{equation}
(corresponding to the orthogonal azimuthal component $U^{\hat\phi}=\Omega/(\gamma r)$), we see that the same equation (41) can alternatively be found by solving the relationship $\Omega^2/(\gamma^2 r^2)=1/n^2$. This relationship means that the negative solution  for the effective potential crosses the $r$ axis in Fig.~3 where the (orthogonal) flow velocity $\Omega/(\gamma r)$ just counterbalances the velocity of light $1/n$ relative to the medium. Inside the radius given by Eq.~(41) the flow velocity exceeds the local velocity of light, and the region is per definition an ergo-region. The physical consequence of this is that light in this region is forced to participate in the positive $\phi$-rotation around the centre, no matter which way the light tries to escape. 

The negative solution of Eq.~(34) has a {\it bottom point}. This point corresponds to unstable circular orbits \cite{leonhardt99}. One might imagine that light, having a value of $b^{-1}$ just corresponding to a turning point at the bottom point, does not go in orbit at all, but stands still. If this scenario were true, light would go directly against the flow at a radial value where the flow velocity equals the velocity of light in the medium.

This idea of still-standing light is however not correct. As was showed above, the point where the flow velocity equals the velocity of light is where the negative solution of the "effective potential" crosses the $r$ axis. The bottom point of this potential is at a larger $r-$ value, and the conclusion that we have to make, is that the bottom point represents an unsteady orbit instead. 

As pointed out by Visser \cite{visser00}, an azimuthal vortex cannot be taken to represent a genuine black hole. This can easily be seen from the "effective potential" plot of a photon possessing positive angular momentum. For, if the photon has a positive $b^{-1}$, its accessible region is of type $R_2$ (cf. also Fig.~2), meaning that the photon ultimately gets reflected out to infinity. The outward turning point can be found arbitrarily close to the vortex singularity. The system has no analogy to a Schwarzschild black hole in the sense of a point of no return existing for all photons.

This does not mean that light cannot be captured by the vortex. As seen from Fig.~3 there is a region beneath the lower curve corresponding to type $R_3$ (cf. again Fig.~2). In this case a photon coming in from infinity is captured by the vortex, and will never escape. This region exists only for light of negative angular momenta (and therefore of negative $b^{-1})$.

Photons having negative angular momenta, thus fighting against the stream, have no outer turning points inside the radius determined by the bottom point of $\hat{b}_-^{-1}$. The photons find the bottom point in some sense to be similar to an event horizon. To be a  genuine horizon, however, the $r-$surface must have the property that no photon emitted on the inside has any possibility to escape, even if emitted only slightly inside and in a radial direction. The bottom-point radius (or the radius in Eq. (41)) does not satisfy this condition, the reason being that the flow has no radial component. Photons can be captured by the vortex, since the flow at an early stage manipulates the trajectories and generates an inward directed spiraling motion for the photons. There is however nothing in the flow itself which drags photons directly towards the centre. If a photon in the inside region is emitted in a straight outward direction, it will continue to increase its $r-$value until it escapes to infinity.

It is instructive in this context to recall the effective potential for Schwarzschild geometry: Photons coming in from infinity and pass the potential's top point, has no outer turning point and will be captured. This does not mean that the radius of the top point corresponds to the event horizon; the Schwarzschild radius lies at a lower $r-$value.

Photons with negative $b^{-1}$ can escape, if the absolute value of $b^{-1}$ is less than the absolute value of the effective potential at the bottom point. Both types of regions, $R_2$ and $R_3$, exist for negative values of $b^{-1}$, as shown in Fig.~3.

\section{Behaviour of light in the general flow}

Combining the two elementary flows considered above, we arrive at a considerably more complex physical picture.  Figure 4 shows the effective potential for the flow when the azimuthal and radial components are given equal weight:  $\Omega^2=\Gamma^2$. It is striking how the previous symmetric perfect-sink diagram from Fig.~2 now becomes skew because of the addition of the vortex flow component.

\begin{figure}[htb]
\centerline{\mbox{\psfig{file=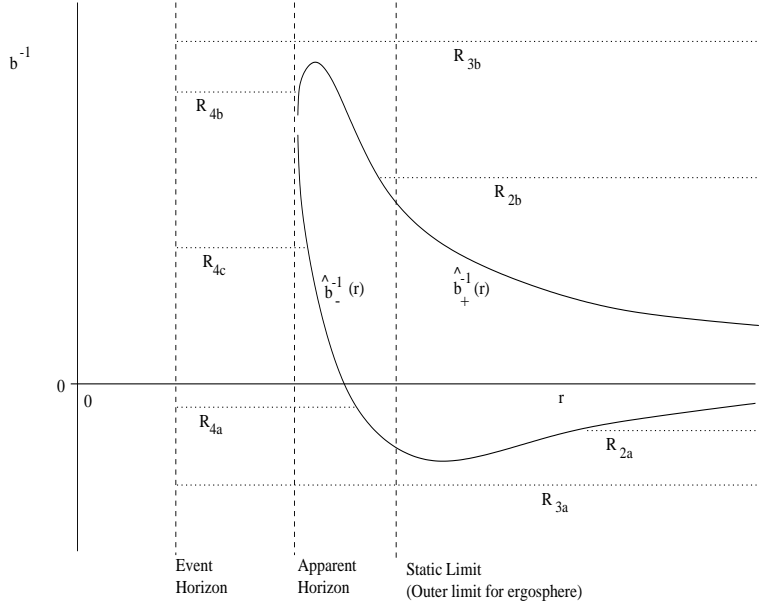,width=10cm,height=8cm,angle=0}}}
\caption{Effective potential, $\hat{b}^{-1}_{\pm}=\hat{E}/\hat{L}$ versus r for light in a medium with both radial and azimuthal flow. Broken lines, $R_2$, $R_3$ and $R_4$ refer to the three possible photon-trajectories.}
\label{krone}
\end{figure}

\subsection{Event horizon}

The event horizon must first be properly defined. In general relativity it is defined as the boundary of the region from which null geodesics can never escape. If the inward-pointing radial component of the fluid velocity exceeds the velocity of light relative to the medium, then, no matter what direction the light ray tries to escape, it will be swept inward by the sink and trapped. This defines the event horizon as the boundary of superluminal radial flow. (An equivalent definition was made by Visser \cite{visser98} for the event horizon in acoustics.)

The mathematical definition of the event horizon is thus
\begin{equation}
\frac{\Gamma^2}{\gamma^2 r^2}=\frac{1}{n^2}
\end{equation}
\label{44}
(cf. Eq.~(10)). Taking into account Eq.~(11) we  get
\begin{equation}
r \equiv r_{event}=\sqrt{\kappa \Gamma^2-\Omega^2}.
\end{equation}
\label{45}
This equation shows that if the strength of the sink for a given vortex is so weak that
\begin{equation}
|\Gamma| < \Omega/\sqrt{\kappa},
\end{equation}
\label{46}
then Eq.~(45) cannot be satisfied. It is noteworthy that the existence of an event horizon thus relies upon the {\it relative} strength of the sink/vortex components, not upon their individual absolute magnitude.

The event horizon is indicated by a vertical line in Fig.~4. A photon existing to the left of this line can never appear on the right hand side.

\subsection{Apparent horizon, ergo region, and static limit}

As already mentioned, a pure sink acts as a static optical black hole and does not distinguish between light having positive or negative angular momenta. Assume now that the sink strength $\Gamma$ is kept constant while the vortex strength $\Omega$ gradually increases. The vortex nature of the flow produces turning points 'higher up' for light having positive $L$ while it 'takes away' turning points having negative $L$. Light propagating along with the azimuthal flow is helped to escape; light fighting against the flow is more easily captured.

When $\Omega$ is further increased the vortex nature gradually takes over, and a plot of the effective potential will be more and more like that of a pure vortex. There remains, however, the following significant difference. When there is a radial flow component, no matter how small in comparison with the azimuthal flow component, the surface where
\begin{equation}
r^2=\kappa \Gamma^2
\end{equation}
\label{47}
will have a special meaning. Inside this surface, which does not depend on $\Omega$ at all, there are no turning points for incoming photons. This surface has much in common with an event horizon (it becomes just the same as the event horizon in the case of a perfect sink; cf. Eq.~(35)), but is in general different from it. We will refer to the surface determined by the condition (47) as the {\it apparent horizon}. This definition of the apparent horizon stems from its visual apparance on the effective potential plot. It must not be confused with the definition made by Visser in \cite{visser98} of the apparent horizon for acoustic waves.

The physical meaning of this concept is the following. Assume, as mentioned, that $\Gamma$ is constant while $\Omega$ gradually increases. Plotting the "effective potential" versus $r$ in the various cases one will see that the azimuthal flow influences the sink flow in much the same way as the rotation of an ordinary black hole influences the black hole's ability to capture light. It is known in general relativity that a rotating black hole captures light less efficiently than a static hole of the same mass \cite{novikov89}. The same kind of effect is found here. If $\Gamma$ is kept fixed while $\Omega$ is increased, the
$b^{-1}$ band of light that will escape increases. This band is the vertical distance between the top- and bottom- point of the effective potential.

The vortex increases the band, but there
is a limit to it. If $\Omega \rightarrow \infty$, the last term in Eq.~(29) becomes negligible and we get
\begin{equation}
\hat{b}_{\pm}^{-1} (\Omega \rightarrow \infty)=\frac{1}{r};
\end{equation}
\label{48}
the positive and negative solutions coincide. Here the meaning of the apparent horizon comes in: from Eq.~(29) we see that the last term becomes imaginary outside $r=r_{app}$, where
\begin{equation}
r_{app}=\sqrt{\kappa}\,|\Gamma|
\end{equation}
\label{49}
is the radius of the apparent horizon. The "effective potential" is thus cut off at the apparent horizon. From Eqs.~(48) and (49) we see that the maximum value of $b^{-1}$ for a photon that can escape from the flow, no matter how large $\Omega$ is chosen, can be defined by
\begin{equation}
b_{esc}^{-1}=\frac{1}{r_{app}}=\frac{1}{\sqrt{\kappa}\,|\Gamma|}.
\end{equation}
\label{50} 
With infinite $\Omega$, the lower curve for the "effective potential" does not cross the abscissa axis, so the maximum interval is $0<b^{-1}<1/r_{app}$.

The apparent horizon is marked by a vertical line in Fig.~4. Geometrically, the apparent horizon is at the $r$-value where the upper curve for the "effective potential" meets the lower, and the curves 'cut each other off'. The term 'apparent' is used, because from a look at the potential plots, this radius appears to be the horizon. The apparent horizon is the point of no return for light coming in from the right on the figure. Inside this horizon there are no turning points, and light entering the region from the outside will never escape.  Recall that the apparent horizon does not depend on $\Omega$, in contrast to the event horizon, Eq.~(45), which depends on both $\Omega$ and $\Gamma$. The event horizon always lies inside the apparent horizon.

The difference between the apparent horizon and the event horizon is experienced by light that is emitted in the inside region. If this region extends beyond the apparent horizon but not beyond the event horizon, it is possible for the ray to escape to infinity. By contrast, in ordinary Schwarzschild geometry, and for the perfect sink, the light pulse must be emitted {\it outside} the apparent horizon if such an escape is to happen, since in these cases the apparent horizon coincides with the event horizon.

The next concept that it becomes natural to introduce for this special kind of flow is the {\it ergo region}. In general relativity, a rotating black hole is known to have an ergo region that reaches outside the event horizon. The part of the ergo region that lies outside the horizon is called the ergosphere. A characteristic property of the ergosphere is that $g_{00}$ becomes negative, so that no state of rest can be defined. Correspondingly, the time Killing vector $\xi_{(t)}$ becomes spacelike, leading in turn to peculiar effects like the Penrose process (cf., for instance, Refs.\cite{novikov89, lightman75}). 

In the present case, certain similarities with a rotating black hole are found. The total flow velocity becomes equal to the velocity of light relative to the medium when
\begin{equation}
\frac{\Omega^2+\Gamma^2}{\gamma^2 r^2}=\frac{1}{n^2}.
\end{equation}
\label{51}
When this equation is satisfied,
\begin{equation}
\tilde{g}_{00}=0.
\end{equation}
\label{52}
For smaller values of $r$, $\tilde{g}_{00}$ becomes negative. The flow thus possesses an effective ergosphere lying inside the position defined by Eq.~(51). 

The ergosphere will always exist for spiral flow. It extends outside the event horizon (if there is one), and also outside the apparent horizon. To verify the last statement, we calculate the total flow velocity:
\begin{equation}
{\bf U}^2=\frac{\Omega^2+\Gamma^2}{\gamma^2 r^2}=\frac{1}{1+r^2/(\Omega^2+\Gamma^2)}.
\end{equation}
\label{53}
At the apparent horizon, we find from this equation that ${\bf U}^2=1/n^2$ if $\Omega=0$ and ${\bf U}^2>1/n^2$ if $\Omega >0$. Thus, at the apparent horizon the total flow velocity is  even for a very small vortex component larger than the velocity of light in the medium. This horizon is therefore located inside the ergosphere.

We will call the outer limit of the ergosphere  the {\it static limit}, in analogy to what is done in general relativity \cite{novikov89}. It will not be further discussed here (although we will briefly return to it in the final section). It is determined by the condition that the absolute flow velocity is equal to the velocity of light relative to the medium, i.e., Eq.~(51). The static limit is marked by a vertical line in Fig.~4.

\subsection{Qualitative features of the light trajectories}

The horizontal lines marked by $R_2 - R_4$ in Fig.~4 correspond to the horizontal lines $R_2 - R_4$ in Fig.~2.

There is no Region 1 in the "effective potential", meaning that there are no stable orbits for light. This is just analogous to the conditions outside a perfect sink, as discussed above.

Region 2 is divided into Regions 2a and 2b in the figure. There is no qualitative difference between the two. They correspond to light entering the flow with an inverse impact parameter that at some point equals the "effective potential". Rays in Region 2  escape to infinity.

Below the potential's bottom point, and above its top point, lie  Regions 3a and 3b. The former represents a photon with negative angular momentum captured by the flow singularity. The latter describes the same situation for a photon with positive angular momentum. It should be noted that Region 3 may also describe a photon that is emitted somewhere outside the event horizon and escapes to infinity.

Region 4 refers to photons emitted somewhere on the inside of the top- or bottom points of the "effective potential". Note that the $R_4$- lines cross the apparent horizon; there exists no condition that forbid photons to cross the apparent horizon from the inside. The event horizon, however, forbids any crossing so that the horizontal lines end there.

Region 4 is divided into three subregions. Region $R_{4a}$ lies on the negative half of the the $b^{-1}$ axis, whereas $R_{4b}$ and $R_{4c}$ lie on the positive half. Two remarks should here be made:

First, Region 4 contains only photons that are emitted in the inner region of the flow and eventually become captured inside the event horizon. Although it remains adequate to keep the quantity $b$ as a formal input parameter, the interpretation of it as an 'impact parameter at infinity' becomes lost. No photons in Region 4 will ever come from, or go to, infinity.

Second, the distinction between the lines $R_{4b}$ and $R_{4c}$ is not physically important. The first line corresponds to a photon with an inner turning point at $\hat{b}_+^{-1}$, while the second line has an inner turning point at $\hat{b}_-^{-1}$. The two curves meet at the point where they are tangential to the apparent - horizon vertical line. Together they enclose the inaccessible region of space, and there is no physical distinction between the two. 

Quite generally, the paths of light will be more complicated here than in Schwarzschild geometry. This is so because of the rotation. What can be read off from our plot of the "effective potential", is which trajectories are captured, and which can escape. In order to calculate the real shape of trajectories, the azimuthal equation of motion has to be solved. This task will not be undertaken in the present paper, but we will discuss the azimuthal equation  briefly in the next section.

\section{Summary, and final remarks}

The usefulness of picturing a non-uniformly moving medium as an effective gravitational field, endowed with a metric $\tilde{g}^{\mu \nu}$, as demonstrated earlier for a pure vertex flow in Ref. \cite{leonhardt00}, turns out to hold true also for the present case with a spiral flow. The effective metric is defined in Eq.~(6); its contravariant and covariant components are given in Eqs.~(20) and (21).

The equation for a photon in Fourier space, Eq.~(4), is in terms of the effective metric picture expressible in the form of the Hamilton-Jacobi equation, Eq.~(5). At the turning point of a photon trajectory, $dr/d\lambda =0$. The turning point solutions of the Hamilton-Jacobi equation are given as solutions of Eq.~(27), in which $\hat{b}^{-1}$ means the turning- point value of the inverse impact parameter $b^{-1}$, in general defined as $b^{-1}=E/L$. There are in general two solutions of the equation, called $\hat{b}_+^{-1}$ and $\hat{b}_-^{-1}$. The solutions are given by Eq.~(29), and are identified as the "effective potential(s)". They are shown in Fig. 2 for pure sink flow, in Fig. 3 for pure vortex flow, and in Fig. 4 for the general spiral case. "Effective potential" considerations turn out to be useful for the qualitative analysis of the various kinds of photon trajectories that can occur, in particular, for the general case of spiral flow.

It is rather remarkable that the "effective potential" possesses properties that are qualitatively similar to those encountered for rotating black holes in general relativity. Thus, we can identify an event horizon, as defined by Eq.~(45), for the flow if its sink component is strong enough in comparison to the vortex component to satisfy the condition $\kappa \Gamma^2 > \Omega^2$. A photon emitted on the inside of the event horizon can never escape to infinity. There exists also an apparent horizon, Eq.~(49), lying outside the event horizon (if the latter exists).   At the apparent horizon, the "effective potential" is cut off. Photons emitted beyond the apparent horizon, but not beyond the event horizon, can escape to infinity. Finally, there is an ergosphere, whose outer boundary is determined by the condition that the flow velocity is equal to the velocity of light relative to the medium, Eq.~(51). In the ergosphere the time Killing vector $\xi_{(t)}$ is spacelike ($\tilde{g}_{00}<0$), so that no state of rest relative to infinity can be defined.

We mention in passing that one interesting possible application of the pure vortex optical black hole is its relationship to the optical Aharonov-Bohm (AB) effect \cite{aharonov59}. This effect is in principle able to explore the long-ranging topological nature of a quantum vortex. The optical AB effect was analysed by Leonhardt and Piwnicki \cite{{leonhardt99},{leonhardt00a}}, assuming a very high value of the refractive index. Because of the AB phase shift around a vortex one may say, strictly speaking, that we are considering angular momenta rather than impact parameters in the present paper. 

We ought finally to make clear what are the limitations of the present kind of theory. First, only nondispersive media are considered. Dispersion - sensitive effects such as the EIT - generated 'slow light' \cite{hau99} are thus not encompassed by the theory. 

Second, although the "effective potential" turns out to be useful for determining the turning points for photons, this concept does not allow us to determine the exact form of the spiral trajectories. If quantitative results are required, then the azimuthal equation of motion has to be solved. This equation can be constructed by considering the $\phi$-component of Eq.~(18):
\begin{equation}
\tilde{k}^{\phi}=\frac{d\phi}{d\lambda},
\end{equation}
\label{54}
and solve the equation
\begin{equation}
\frac{d\phi}{d\lambda}=\tilde{g}^{\phi \nu} k_\nu=\tilde{g}^{\phi 0} k_0+\tilde{g}^{\phi r}k_r+\tilde{g}^{\phi \phi}k_{\phi}.
\end{equation}
\label{55}
If this is done, the actual trajectory can be plotted. A $\phi$-diagram would show special features at the point where the azimuthal flow component becomes of the same magnitude as $c/n$. We have made some calculations in this direction, but the formalism becomes somewhat unwieldy and is not reproduced here.

\subsection{On the physical meaning of the horizon}

It is of interest to carry out the discussion of the physical meaning of the horizon a bit further. In general relativity, a horizon has interesting quantum properties that have their roots in the behaviour of classical waves. This is related to the phase of a wave developing a logarithmic singularity at the horizon. The well known Hawking radiation stems from a Bogoliubov transformation relating the "in" vacuum to the "out" vacuum  \cite{birrell84}. One may wonder: does our physical system allow Hawking radiation to occur?

To investigate this point further, let us first recall some facts about the behaviour of a scalar field $\Phi$ near the horizon $r=2M$ of a Schwarzschild black hole (for simplicity we now put $G=c=1$). As shown by Jensen and Candelas  \cite{jensen86}, near the horizon the solution of the scalar field equation  $ \square  \Phi =0$,
 regular at infinity, varies with $r$ as
\begin{equation}
 \Phi \propto \left( \frac{r}{2M}-1 \right)^{-2iM \omega }.       
\end{equation}
\label{56}
The wavelengths become shortened near the horizon; the waves pile up. The effect is demonstrated graphically, for instance in Ref. \cite{brevik01} (dealing with the somewhat more general case of a Schwarzschild - de Sitter scalar field). It is worth noticing here that the effect is quite analogous to that taking place near the horizon in a Rindler frame. In that case, the field mode corresponding to Eq.~(56) is found to vary as the modified Bessel function $K_{i\omega}$ of imaginary order \cite{takagi86}
\begin{equation}
 \Phi \propto K_{i\omega}(kx). 
\end{equation}
\label{57}
Here $k=|{\bf k}_\perp | $ is the magnitude of the transverse wave vector ${\bf k}_\perp $, and $x$ is the longitudinal distance from the horizon. Since $K_{i\omega}(kx)=\frac{1}{2}\Gamma (i\omega )(\frac{1}{2}kx)^{-i\omega }$ for small arguments, it follows that $\Phi \propto x^{-i\omega}$, thus in essential agreement with Eq.~(56).

After these preliminaries we return to our flow field. To avoid complicating the formalism too much, we imagine a scalar field $\Phi$ that is azimuthally symmetric. In the effective metric picture the field equation is 
\begin{equation}
 \partial_\mu \left( \sqrt{-\tilde{g}}\tilde{g}^{\mu \nu}\partial_\nu \Phi \right) =0 ,
\end{equation}
\label{58}
where the indices $\mu$ and $\nu$ run over 0 and $r$. From Eqs.~(21) we calculate the metric
 \[ \tilde{g}=-\left[ 1-\frac{\kappa}{n^2}\left( 1+\frac{\Omega^2+\Gamma^2}{r^2} \right) \right]\left(r^2+\frac{\kappa (\Omega^2+\Gamma^2)}{n^2}\right) \]
\begin{equation}
-\frac{\kappa^2(\Omega^2+\Gamma^2)}{n^4}\left( 1+\frac{\Omega^2+\Gamma^2}{r^2}\right).
\end{equation}
\label{59}
We will investigate whether there are solutions of Eq.~(59) that are singular at the horizon. As "horizon" it is natural here first to consider the static limit, $r=r_{stat. limit}$, since $\tilde{g}_{00}=0$ there. According to Eqs.~(51) or (21),
\begin{equation}
r_{stat. limit}=\sqrt{\kappa (\Omega^2+\Gamma^2)}.
\end{equation}
\label{60}
Consider now positions $r$ just outside the static limit. As the metric approaches a constant value in this limit, $\tilde{g} \rightarrow -(\kappa /n^2)(\Omega^2+\Gamma^2)$, we can omit the influence from this factor altogether when looking for singular solutions, and consider instead the reduced equation
\begin{equation}
 \partial_\mu \left( \tilde{g}^{\mu\nu}\partial_\nu \Phi \right)=0.
\end{equation}
\label{61} 
Assuming a time factor $\exp(i\omega t)$ we can write this equation approximatively as
\[ \left( n^2+\frac{\kappa (\Omega^2+\Gamma^2)}{r^2}\right) \omega^2 \Phi-\frac{2i\omega\kappa \Gamma}{r}\sqrt{1+\frac{\Omega^2+\Gamma^2}{r^2}}\,\partial_r\Phi \]
\begin{equation}
+\partial_r \left[ \left(1-\frac{\kappa \Gamma^2}{r^2}\right)\partial_r\Phi \right]=0.
\end{equation}
\label{62}
This equation actually has no solution singular at the static limit. Such a solution occurs, however, at the {\it apparent horizon}, $r=r_{app}=\sqrt{\kappa}\,|\Gamma |$. The first term in Eq.~(62) can be neglected to leading order, and the scalar mode vanishing at $r \rightarrow \infty $ becomes, close to the apparent horizon,
\begin{equation}
 \Phi \propto \left( 1-\frac{\kappa \Gamma^2}{r^2}\right)^{-i\omega \sqrt{\kappa (\Omega^2+n^2\Gamma^2)} }.
\end{equation}
\label{63}
This field, in fact, has the same qualitative behaviour as that outside a Schwarzschild black hole, Eq.~(56). Physically, this singular behaviour can be understood from the fact that the apparent horizon is the point of no return for light coming in from infinity.

Does this imply that there really occurs Hawking radiation from the optical black hole? The answer is in our opinion no, the reason being  that the metric $\tilde{g}_{\mu\nu}$ is an artificial metric only.  Our fluid system is physically a special-relativistic system. There exists  one symmetry group in the present case, namely the Poincar{\'e} group, according to which the vector $\partial/\partial t$ is a Killing vector of spacetime (cf. page 45 in \cite{birrell84}). The condition for Hawking radiation to occur, is that there are physically inequivalent "in" and "out" vacua. That is the case in {\it curved} spacetime, where the Poincar{\'e} group is no longer a symmetry group. (This point is however not quite trivial, and it is only fair to mention that Hawking radiation has been claimed to occur in recent discussions on acoustic black holes.)

\end{document}